\documentclass[prb,showpacs,preprintnumbers,preprint,superscriptaddress]{revtex4}

\usepackage{graphicx}
\usepackage{dcolumn}
\usepackage{bm}
\usepackage{amssymb}
\usepackage{color}

\begin{document}

\title{Cloaking dielectric spherical objects by a shell of metallic nanoparticles}

\author{Stefan M\"uhlig}
 \affiliation{Institute of Condensed Matter Theory and Solid State Optics, Abbe Center of Photonics,
 Friedrich-Schiller-Universit\"at Jena, D-07743 Jena, Germany}

\author{Mohamed Farhat}
 \affiliation{Institute of Condensed Matter Theory and Solid State Optics, Abbe Center of Photonics,
 Friedrich-Schiller-Universit\"at Jena, D-07743 Jena, Germany}

\author{Carsten Rockstuhl}
 \affiliation{Institute of Condensed Matter Theory and Solid State Optics, Abbe Center of Photonics,
 Friedrich-Schiller-Universit\"at Jena, D-07743 Jena, Germany}

\author{Falk Lederer}
 \affiliation{Institute of Condensed Matter Theory and Solid State Optics, Abbe Center of Photonics,
 Friedrich-Schiller-Universit\"at Jena, D-07743 Jena, Germany}

\date{\today}

\begin{abstract}
We show that dielectric spheres can be cloaked by a shell of amorphously arranged metallic nanoparticles. The shell represents an artificial medium with tunable effective properties that can be adjusted such that the scattered signals of shell and sphere almost cancel each other.  We provide an analytical model for the cloak design and prove numerically that the cloak operates as desired. We show that more than  $70 \%$ of the scattered signal of the sphere can be suppressed at the design wavelength. Advantages and disadvantages of such a cloak when compared to other implementations are disclosed.
\end{abstract}

\pacs{42.25.Bs, 41.20.-q, 42.79.-e}
\maketitle

\section{Introduction}

In $2006$, Pendry \textit{et al.} have shown that by embedding a finite size object into a coating made of a suitable metamaterial (MM), this object can be concealed from an external observer; i.e. it can be cloaked \cite{pendry1-2006}. The cornerstone of this work was the understanding that a geometrical transformation  can be cast into a suitable spatial distribution of a biaxial material. The transformation of the cloak itself is an inflation of space such that a point acquires the finite dimension of a sphere or cylinder in which objects can be hidden. At first glance it seemed to be detrimental that an efficient cloak requires both controllable (artificial)  permittivity and permeability. For the  latter is inaccessible in natural media their implementation would require MMs. MMs are artificial media with optical properties that can be controlled by suitably designed unit cells. It was soon understood that although a perfect cloak requires quite arbitrary permittivity and permeability, simplified cloaks can be implemented which only need the control of either
 quantity. Such simplified cloaks may similarly guide light around the object to be cloaked, however,  at the expense that  impedance matching to the surrounding is incomplete  and spurious reflections occur \cite{pendry2-2006}. A team led by Pendry and Smith finally implemented a simplified cloak using a MM consisting of concentric layers of split-ring resonators. The cloak made a copper cylinder invisible to an incident plane wave at 8.5 GHz \cite{pendry2-2006}. Another implementation of such a cloak was suggested by relying on elliptical metallic nanoparticles (NPs) \cite{shalaev-cloak}. There the ellipticity varied across the radius to evoke the desired radial and azimuthal permittivity profile across the cylindrical cloak. This can only be achieved if the material forming the cloak may be considered as an effectively homogenous medium with tunable properties.

Farhat \textit{et al.} \cite{farhatoe} analyzed cloaking of transverse electric (TE) fields through homogenization of radially symmetric metallic structures. The cloak consisted of concentric layers cut into a large number of small infinitely conducting sectors which is equivalent to a highly anisotropic permittivity. This structure was shown to work for different wavelengths provided that they are ten times larger than the size of the unit cells building the cloak.

An alternative approach to design a cloak was put forward by Leonhardt which relies on conformal transformation \cite{ulf,ulfnjp}. This approach is equally versatile and many other optical devices next to a cloak can be envisioned using such a design strategy \cite{Schmiele}. An even further different path to invisibility was suggested by McPhedran \textit{et al.} \cite{ross-94,rosscloak}. They proposed to cloak a countable set of line sources when they are situated near to a cylindrical coating filled with a negative index material using anomalous resonance. More recently, Tretyakov \textit{et al.} \cite{tretyakov} have demonstrated the possibility of designing a cloak for microwaves by using simple structures made of metallic layers. This technique permits broadband and low-loss cloaking as experimentally demonstrated .

In 2005, Al\`u and Engheta proposed the use of plasmonic materials to render small dielectric or conducting objects nearly invisible. There the mechanism relies on a scattering cancelation technique based on
the negative local polarizability of a cover made of a metallic
material \cite{alu-pre}.
This cloak has been shown to be relatively robust against
changes in the object geometry and operation wavelength.
Recently, its first experimental realization and characterization at
microwave frequencies has been performed \cite{alu-exp}.
Applications for stealth technology, non-invasive probing and sensing can be envisaged,
opening the door to various applications in medicine, defense and telecommunications. As could be already seen from this first experimental implementation, the spectral range were such a cloak can be implemented can be increased when the material of the shell itself is an artificial medium rather than a homogenous metal. This relaxes design constraints and allows, moreover, for the potential exploitation of new fabrication methods to realize such a cloak.

In this contribution we propose and analyze in depth a cloak where the plasmonic shell surrounding the object to be concealed is already an artificial medium with \emph{tunable} effective properties. To this end we focus on a plasmonic shell that is composed of a large number of metallic NPs which has already been used for building metamaterial nanotips \cite{rockstuhlapl} or bottom-up bulk metamaterials \cite{muhlig}. The fabrication of such a shell of metallic NPs is amenable by self-organization methods in colloidal nano-chemistry \cite{burgi}. We analytically demonstrate but also verify by full-wave simulations that such a shell allows to reduce the scattering cross-section of the dielectric core to be concealed by orders of magnitude. The underlying mechanism that enables the design of the cloak is the treatment of the shell as an effective medium which shows, according to Maxwell-Garnett theory, a strong dispersion in its effective permittivity around the frequency were a localized plasmon polariton is excited in the metallic NPs. The associated very large, very small, or even close to zero values for the effective permittivity can be exploited to cancel the scattering response from the core at tunable frequencies. In our work we combine a simplified description that treats the problem analytically in the quasi-static limit with rigorous simulations where all the particles involved are explicitly considered. The advantages and disadvantages of such a cloak are discussed in detail in this contribution. Moreover, besides the mere exploration of such a cloaking application we show here that effective properties assigned to MMs can be further exploited in the design of a functional device.  The ability to compare predictions of optical properties of such finite systems obtained from a simplified description to full-wave simulations provides confidence that MMs can be faithfully considered in the future as ingredients to design bulk and macroscopic optical devices and applications.  This will be feasible by looking effective MM properties up from tables as it is now possible for natural materials, just as we do here in designing the desired structure. The specific fine structure of the MMs does not need to be considered in this process.

\section{Formulation of the problem}

By considering the scattering of an illuminating electromagnetic plane wave
at a given object (centered without loss of generality at the origin of a spherical coordinate system), it can be shown \cite{jackson} that the scattered electric
and magnetic fields ${\bf E}_\mathrm{s}$ and ${\bf H}_\mathrm{s}$ can be expressed in terms
of the coefficients $c_{nm}^{\mathrm{TE}}$ and $c_{nm}^{\mathrm{TM}}$. These coefficients represent the amplitudes of the spherical harmonics  into which the scattered field can always be expanded where TE and TM refer to the electric and magnetic contribution, respectively. The complex amplitudes depend on the geometry of the scatterer, the material it is made of and on the frequency. Moreover, for a given size $a$ of a spherical scatterer, only amplitudes with $n<N$ are relevant since these scattering amplitudes are of the order of $(k_0 a)^{(2n+1)}$ where $k_0=\omega/c\sqrt{\varepsilon_\mathrm{b}\mu_\mathrm{b}}$ is the wave number of the background region, and $\varepsilon_\mathrm{b}\, ,\mu_\mathrm{b}$ are its permittivity and permeability. For spherical objects and for an incident plane wave propagating along $z$ direction and being polarized parallelly to $x$ direction, the high-symmetry of the object suggests that only coefficients with $n=m$ have non-zero-amplitudes.
These coefficients of order $(n)$ can be calculated for a sphere as
\begin{equation}
c_{n}^{\mathrm{TE}}=-\frac{U_{n}^{\mathrm{TE}}}{U_{n}^{\mathrm{TE}}+V_{n}^{\mathrm{TE}}}\, , \qquad
c_{n}^{\mathrm{TM}}=-\frac{U_{n}^{\mathrm{TM}}}{U_{n}^{\mathrm{TM}}+V_{n}^{\mathrm{TM}}}
\label{coef}
\end{equation}
where $U_{n}^{\mathrm{TE,TM}}$ and $V_{n}^{\mathrm{TE,TM}}$ are determinants of matrices comprising spherical Bessel functions (see \cite{alu-pre}). They can be explicitly calculated once all the parameters describing the system are fixed.

Now, if an observer is placed in close proximity to the scatterer (near-field) or far from it (far-field), the possibility of detecting the scatterer's presence is entirely given by the total scattering cross section (SCS) defined as \cite{jackson}
\begin{equation}
\sigma_s=\frac{2\pi}{\left|k_0\right|^2}\sum_{n=1}^{\infty}(2n+1)
\bigg\{\left|c_{n}^{\mathrm{TE}}\right|^2+\left|c_{n}^{\mathrm{TM}}\right|^2\bigg\}.
\label{scs}
\end{equation}
Therefore, suppressing this SCS to the largest possible extent conceals an object.

\begin{figure}[h!]
\begin{center}
\scalebox{0.55}{\includegraphics{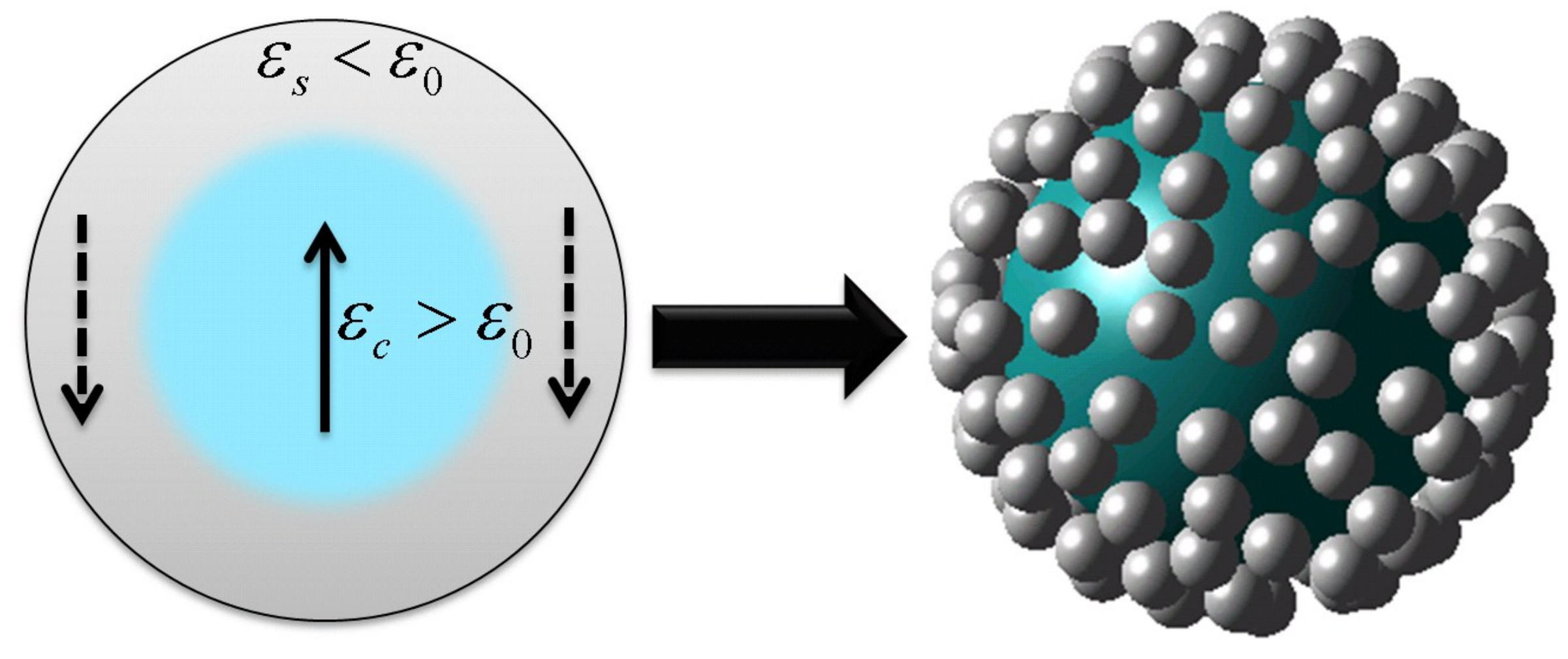}}
\caption{Schematic of the dielectric sphere to be cloaked surrounded by metallic nanoparticles (right) forming an effective invisibility shell (left) described by its effective permittivity ${\varepsilon_\mathrm{s}}$ and effective polarization vector (dashed arrows) ${\bf P_\mathrm{s}=\varepsilon_0(\varepsilon_\mathrm{s}-1){\bf E}}$ of opposite direction and same amplitude as the vector of the bare object (solid arrow) ${\bf P_\mathrm{c}=\varepsilon_0(\varepsilon_\mathrm{c}-1){\bf E}}$, where ${\bf E}$ is the local electric field and where our background material is air.}
\label{scheme}
\end{center}
\end{figure}

By considering a scatterer sufficiently small with respect to the wavelength of interest, it can be shown that the SCS is only dominated by the lowest TE scattering multipole (of order $n~=~1$).
In this case, it was shown in literature \cite{alu-pre,alu-exp}, that by covering the object with a material with low or negative permittivity, it is possible to significantly reduce the SCS at certain frequencies.
This effect may be explained by considering the first order electric multipole ($n~=~1$) which is related to an integral of the polarization vector
%${\bf P}_1(\mathbf{r},\omega)=\left[\varepsilon_1(\mathbf{r},\omega)-\varepsilon_0\right]{\bf E}_1(\mathbf{r},\omega)$
${\bf P}_1(\mathbf{r},\omega)=\varepsilon_0\left[\varepsilon(\mathbf{r},\omega)-\varepsilon_{\mathrm{b}}\right]{\bf E}(\mathbf{r},\omega)$ inside the material \cite{jackson}, where $\varepsilon_1(\mathbf{r},\omega)$ and ${\bf E}(\mathbf{r},\omega)$ are the local relative permittivity and electric field, respectively. As can be seen from Fig. \ref{scheme} a plasmonic cover with less than one permittivity induces a local $\pi$ out-of-phase polarization vector with respect to the local electric field, thus permitting partial or even entire cancelation of the scattering signal caused by the object.

In the quasi-static limit where the size of the dielectric core sphere is much less than the wavelength and only the lowest order Mie coefficient is important, an analytical result can be provided that links the core radius $a_\mathrm{c}$ and its permittivity $\varepsilon_\mathrm{c}$ to the required shell radius $a_\mathrm{s}$ and permittivity $\varepsilon_\mathrm{s}(\omega)$ in order to suppress the entire scattered signal. It reads as \cite{alu-pre}
\begin{equation}
\gamma^3=\frac{\left[\varepsilon_\mathrm{s}(\omega)-\varepsilon_\mathrm{b}\right]\left[2\varepsilon_\mathrm{s}(\omega)+\varepsilon_\mathrm{c}\right]}
{\left[\varepsilon_\mathrm{s}(\omega)-\varepsilon_\mathrm{c}\right]\left[2\varepsilon_\mathrm{s}(\omega)+\varepsilon_\mathrm{b}\right]}
\label{cloaking_qs}
\end{equation}
with $\gamma=a_\mathrm{c}/a_\mathrm{s}$ and $\varepsilon_\mathrm{b}$ being the permittivity of the surrounding. A shell, in simple words, that possesses the properties specified by this equation acts as an anti-reflection coating for the core.

\begin{figure}[h!]
\begin{center}
{\bf (a)} \scalebox{0.475}{\includegraphics{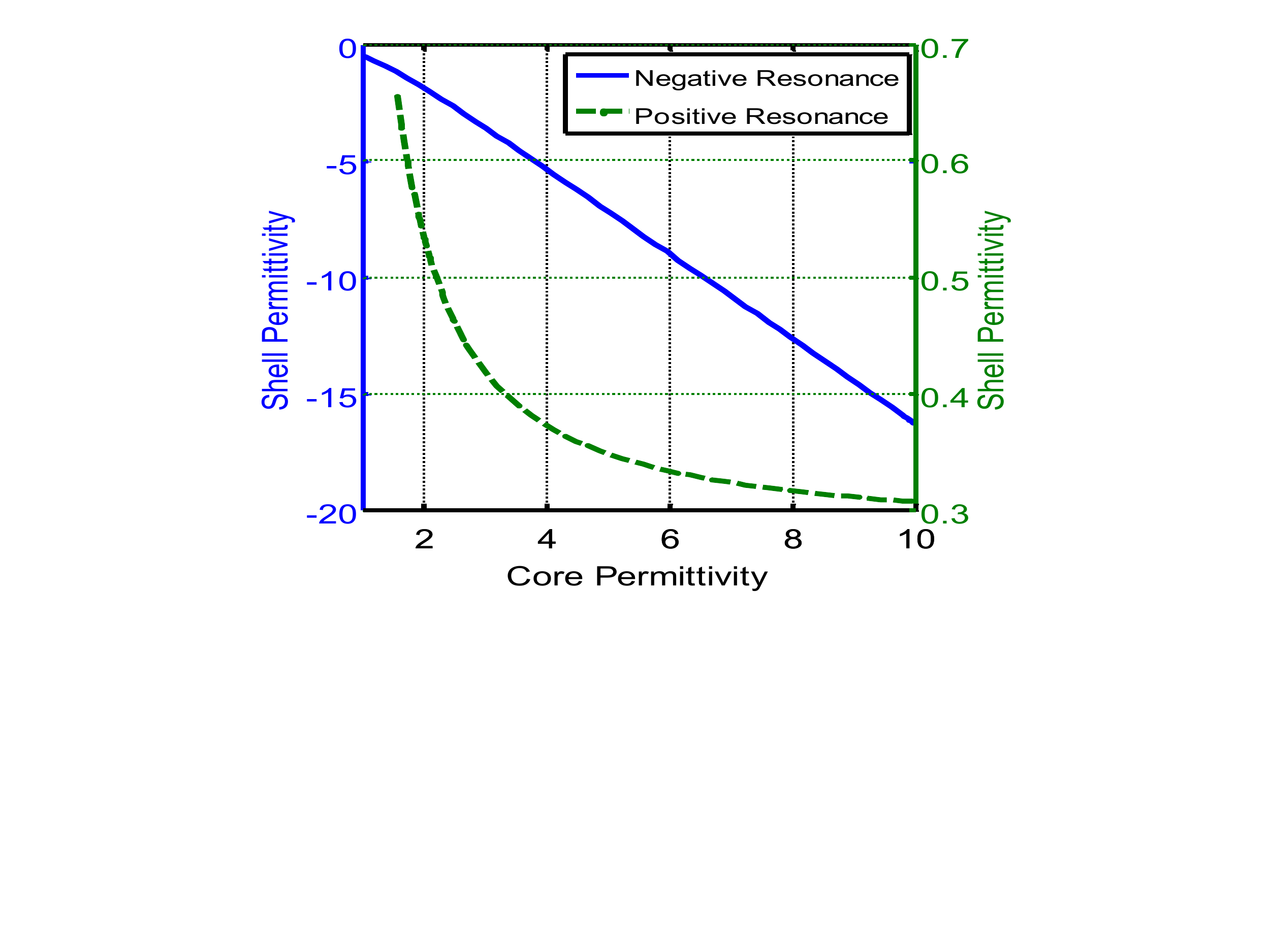}}
%\vspace{1cm}
{\bf (b)} \scalebox{0.475}{\includegraphics{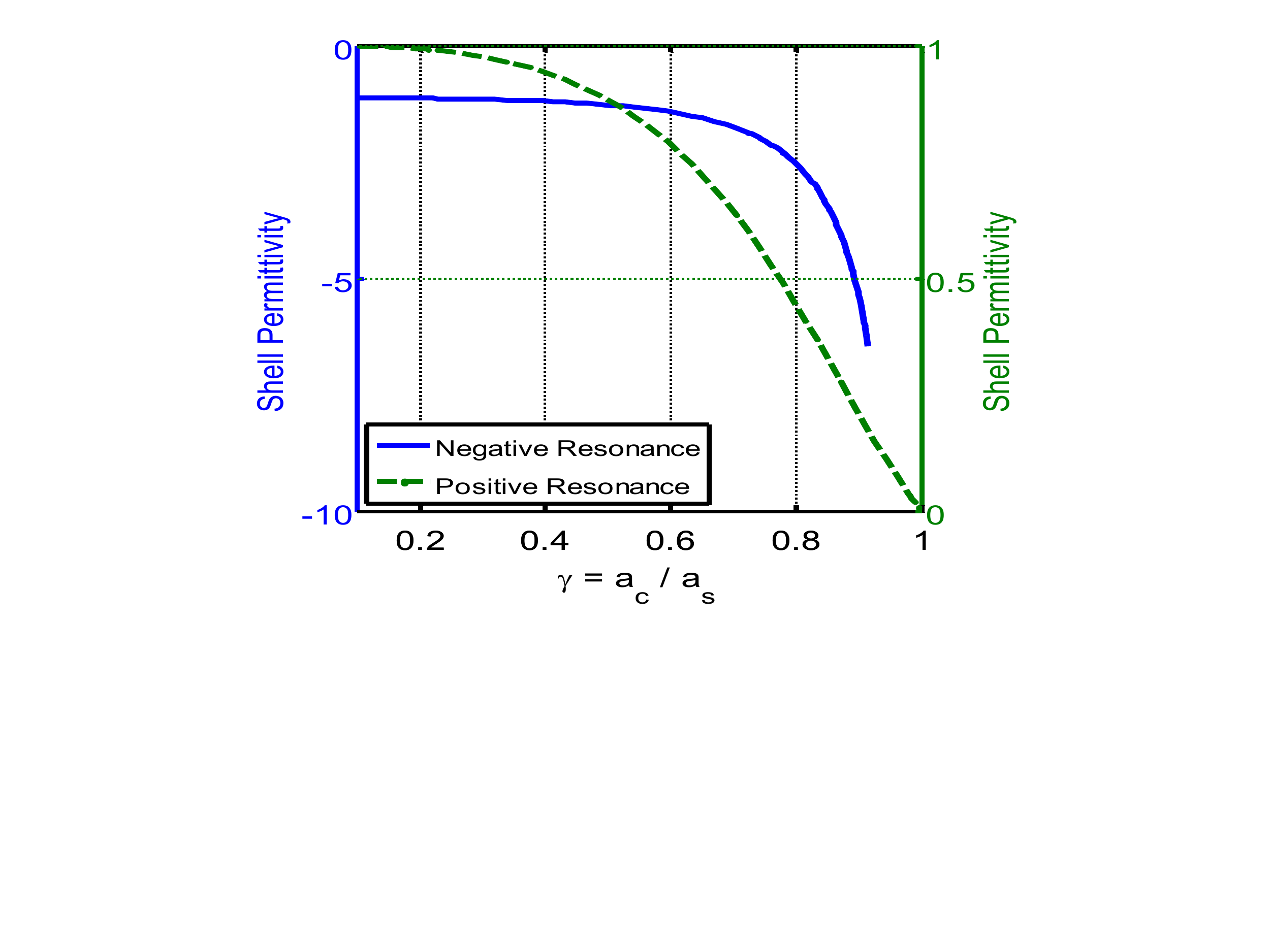}}
\caption{(a) Shell permittivity for both negative (blue solid line) and positive (green dashed line) solutions of Eq. \ref{cloaking_qs} that allows to cloak a sphere with a radius of 35 nm as a function of its permittivity with a shell of a thickness of 10 nm. The permittivity of the surrounding was set to be 1. (b) Required shell permittivity ${\varepsilon_\mathrm{s}}$ as a function of the ratio core to shell radius for negative (blue solid line) and positive (green dashed line) solutions by assuming that the core has a permittivity of ${\varepsilon_\mathrm{c}}= 2.25$ and the surrounding medium again has a permittivity of 1.}
\label{sweep}
\end{center}
\end{figure}

In this paper, we will focus on a cloak for small dielectric spheres (approximately 10 times smaller than the optical wavelength , i.e., in the order of 50 nm.) which requires an effective shell of definite radius and permittivity in order to reduce the scattering by annihilating its dipole moment. For a given core permittivity $\varepsilon_\mathrm{c}$ there are two different solutions to Eq.~\ref{cloaking_qs}, corresponding to a negative and a positive permittivity for the shell. In Fig.~\ref{sweep} both solutions for the required shell permittivity to cloak the core are shown as function of the core permittivity. For the positive solution we notice that the larger the core permittivity the smaller the required shell permittivity $\varepsilon_\mathrm{s}$. By inspecting the negative solution we find that a larger core permittivity implies a more negative shell permittivity. For a fixed value of $\varepsilon_\mathrm{c}=2.25$ we have furthermore solved Eq.~\ref{cloaking_qs} and plotted $\varepsilon_\mathrm{s}$ against the ratio of the shell to the core radius $\gamma$. It suggests that for a rather small ratio, $\varepsilon_\mathrm{s}\approx 1$ is enough to drastically reduce the scattering (which is obvious because of the low visibility of the small sphere if $\gamma<<1$).

We are interested in realistic structures where the shell will be implemented by metallic nanospheres that are limited in size for experimental reasons (a thin cloaking shell is more convenient for experimental applications). Therefore, a sufficiently small  effective permittivity of the shell is required. The cloak is designed to operate in a regime where $\gamma$ is only slightly smaller than unity.
To get specific with regard to the sphere to be cloaked we fix the core sphere radius to $a_\mathrm{c}~=~35$ nm and vary only the core permittivity. Furthermore the radius of the core-shell system is set to  $a_\mathrm{s}~=~45$ nm.
Since the shell will be implemented by metallic NPs, their uniform diameter amounts to 10 nm. The system we consider is sketched in Fig.~\ref{scheme}. The shell formed by the metallic NPs mimics a homogenous medium to which effective properties can be assigned. At a certain design frequency the effective properties of this medium have to be chosen such that Eq.~\ref{cloaking_qs} is fulfilled. In the succeeding section effective parameters will be assigned to the shell in the framework of Maxwell-Garnett theory. Such a treatment allows furthermore for an analytical design of the cloak in the quasi-static limit. We are going to verify the functionality of the cloak by modeling the entire structure as shown in Fig.~\ref{scheme} by full-wave simulations and to compare the results to those of the analytical consideration.

\section{Effective medium approach - Maxwell-Garnett approximation}

\begin{figure}[h!]
\begin{center}
{\bf (a)} \scalebox{0.475}{\includegraphics{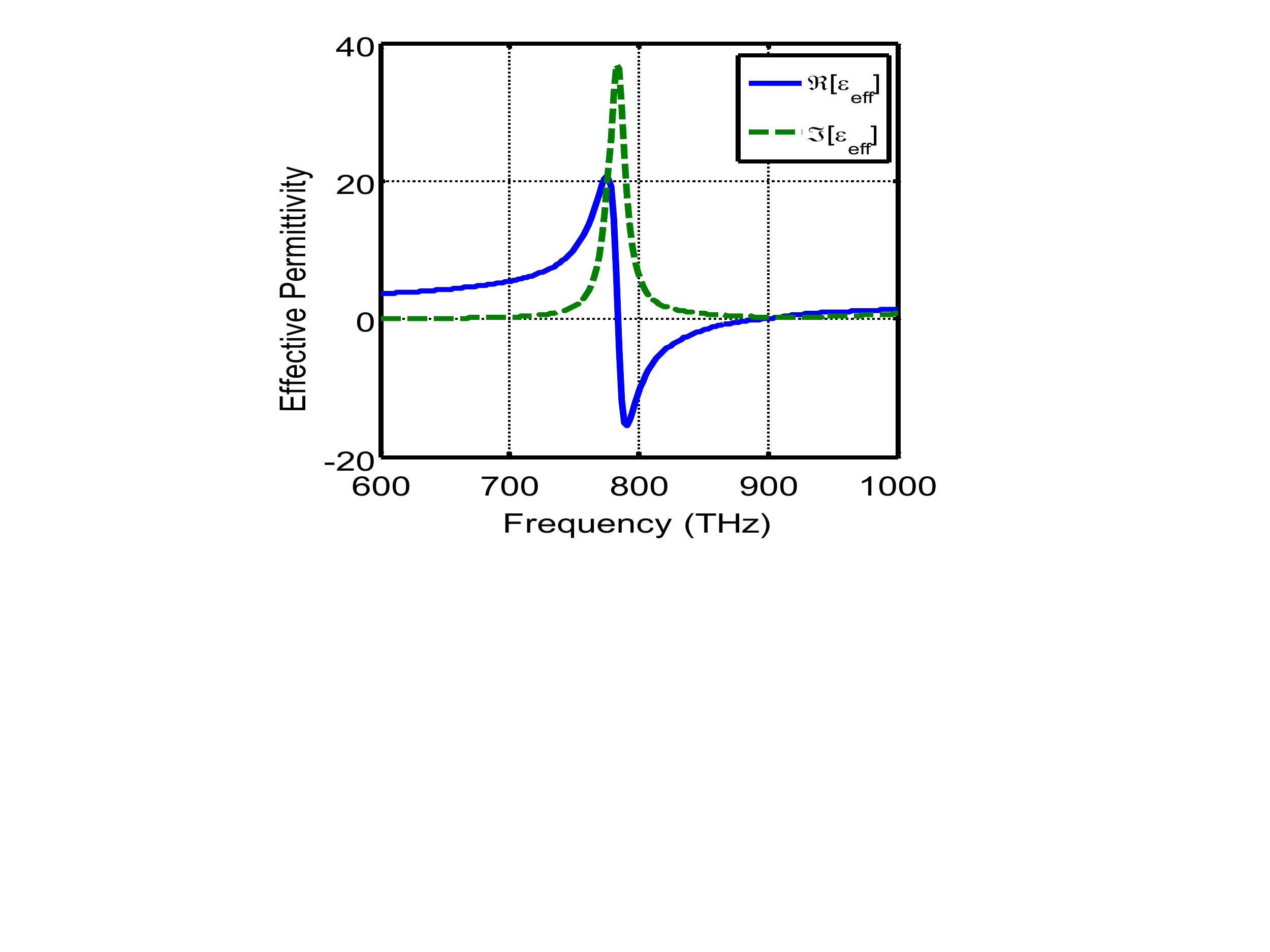}}
%\vspace{1cm}
{\bf (b)} \scalebox{0.475}{\includegraphics{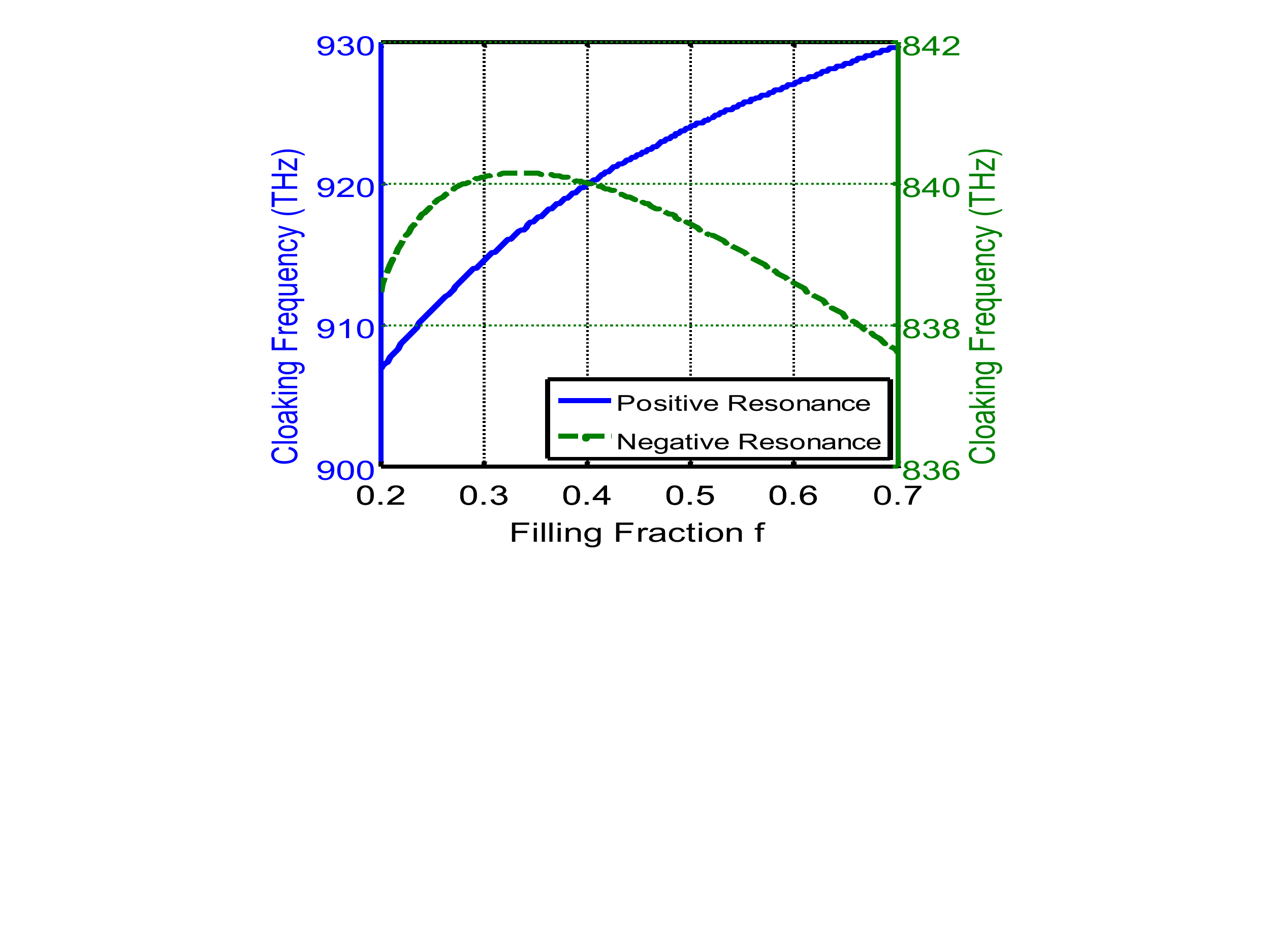}}
\caption{(a) Effective real (blue solid line) and imaginary (green dashed line) parts of the permittivity of the shell given by Maxwell-Garnett theory which is valid for low filling fractions (here we have a filling fraction of 0.34). (b) Cloaking frequency versus the filling fraction of the shell for both positive (blue solid line) and negative (green dashed line) cloaking.}
\label{mg}
\end{center}
\end{figure}

In this section we will analyze analytically the properties of the MMs made of metal NPs. In fact, the description of the shell material as an effective medium is a reasonable approximation for small metallic NPs when compared to the wavelength of light. Moreover, here the NPs are distributed at the surface of the core in an amorphous manner with a low filling fraction (around 0.34). Hence, all all of these properties justify the application of an effective medium approach where we use the Maxwell-Garnett approximation. The spheres were assumed to be made of silver where the tabulated permittivity \cite{christy} has been used in the simulations .

The shell can be described by an effective permittivity which is given by the Maxwell-Garnett formula as (see e.g. \cite{sihvola})
\begin{equation}
\varepsilon_{\mathrm{eff}}(\omega)=\varepsilon_{\mathrm{m}}\frac{\varepsilon_\mathrm{i}(\omega)
[1+2f]-\varepsilon_\mathrm{m}[2f-2]}{\varepsilon_\mathrm{m}[2+f]+\varepsilon_\mathrm{i}(\omega)[1-f]}
\label{mg_eq}
\end{equation}
where $\varepsilon_\mathrm{m}$ is the non-dispersive permittivity of the host medium (air in our case), $\varepsilon_\mathrm{i}(\omega)$ is the one of the NP and $f$ is the filling fraction of the effective medium.

Figure~\ref{mg} (a) shows the real and imaginary part of the permittivity for such a medium with the filling fraction of 0.34. It can be seen that the effective permittivity of the MM exhibits a strong dispersion with a Lorentzian line shape near the frequency where the localized plasmon polariton is excited in the metallic NP. For higher filling fractions $f$, this dispersion curve is shifted towards smaller frequencies. This can be explained by the mutual coupling of the resonant fields in adjacent metallic spheres. It has to be mentioned that the resonance can be tuned across a much larger spectral domain if small metallic shells would have been considered rather than homogenous spheres. Moreover, it can be seen in Fig.~\ref{mg} (a) that the MM behaves differently below and above the resonance frequency. At larger frequencies it mimics a metal [$\Re(\varepsilon_{\mathrm{\mathrm{eff}}})~<~0$] whereas at smaller ones it behaves as a dielectric material [$\Re(\varepsilon_{\mathrm{\mathrm{eff}}})~>~0$). According to the solution of Eq.~\ref{cloaking_qs}, the regime of interest to design a cloak for the core sphere is either the dielectric one far off-resonance where $\varepsilon_{\mathrm{eff}}$ is less than one but larger than zero. In addition we require moderate losses at the frequency of operation [$\Im(\varepsilon_{\mathrm{eff}})~<<~\Re(\varepsilon_{\mathrm{eff}})$].
Both requirements are suitably matched for the given filling fraction at a frequency slightly exceeding 900 THz. Moreover, generally we can use also an operational wavelength where the other condition for the shell is met, i.e., exhibiting a large negative permittivity. However, in this case the MM has to be operated near the resonance  where the strong absorption will greatly deteriorate its performance. Moreover, closer to the resonance the validity of the effective medium theory becomes questionable. Both aspects will be investigated in detail once the performance of the cloak is analyzed with full-wave simulations. In addition, from such simulations the anticipated operational frequencies of the cloak can be extracted in a unique manner.

Since we have fixed the NP size, the only remaining degree of freedom in the design of the cloak is the filling fraction. It sensitively affects the effective properties of the MM shell, most notably by scaling the height of the resonance oscillator strength. This changes the frequency where the cloak shall operate. In fact,  this can be easily seen by inserting Eq. \ref{mg_eq} into Eq. \ref{cloaking_qs} where we get the required filling fraction for scattering cancellation. In Fig~ \ref{mg} (b) we show the operational frequency where the effective medium of the shell takes the required values as a function of the filling fraction. Both, the positive and the negative solution are shown. With the parameters given above we require here as solution  $\varepsilon_{\mathrm{eff}} (\omega)~=~0.49$ (positive solution) or $\varepsilon_{\mathrm{eff}}(\omega)~=~-2.28$ (negative solution), respectively.

From this family of solutions we have chosen those which correspond to a filling fraction of  $f$~=~0.34. This basically fixes the operational frequency. Moreover, this very filling fraction seems to be reasonable since the Maxwell-Garnett approximation still holds \cite{sihvola}. In the following we verify the functionality of the cloak using full-wave simulations.

\section{Numerical analysis of the cloaking effect}

\begin{figure}[h!]
\begin{center}
{\bf (a)} \scalebox{0.5}{\includegraphics{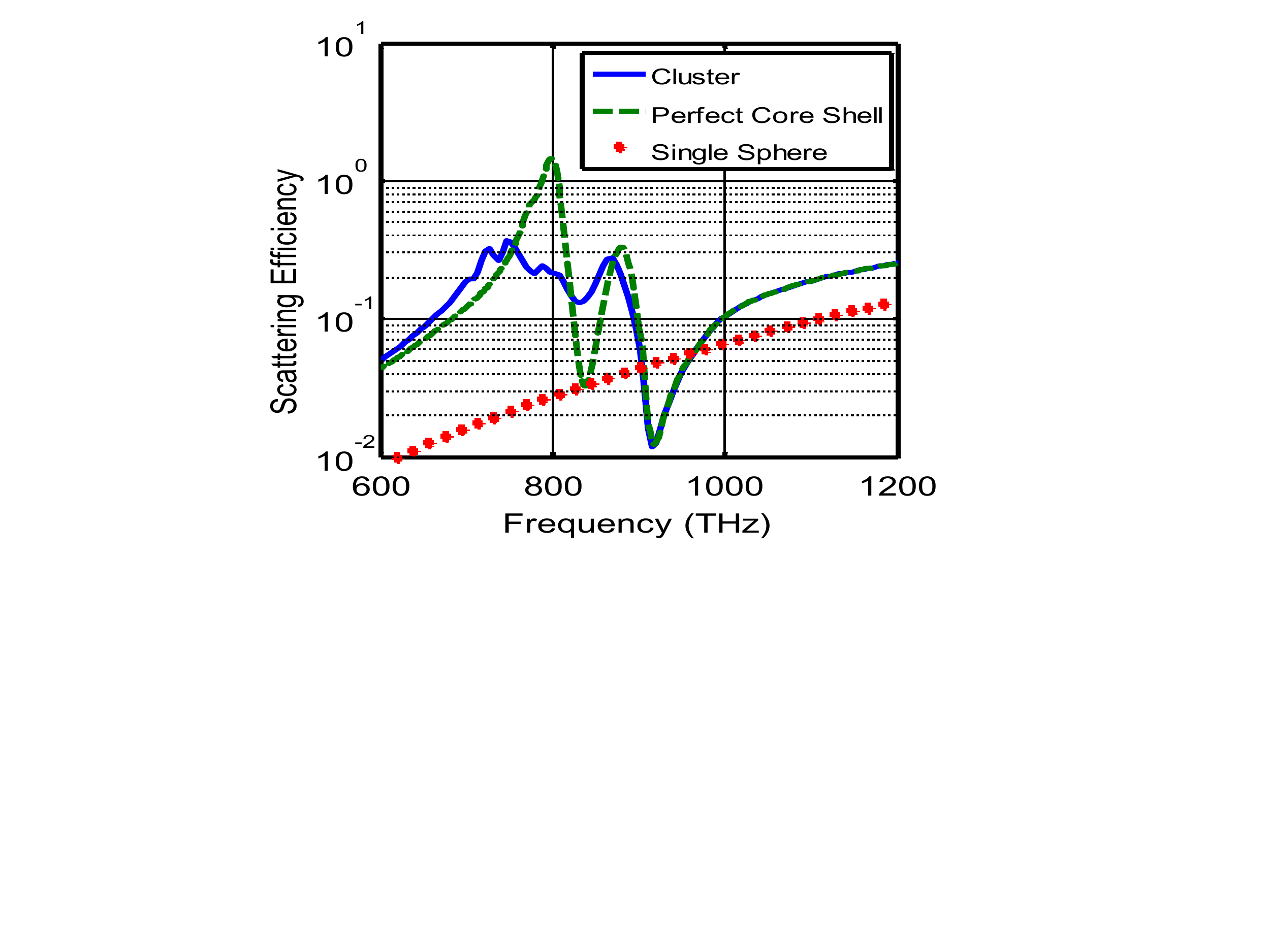}}
{\bf (b)} \scalebox{0.5}{\includegraphics{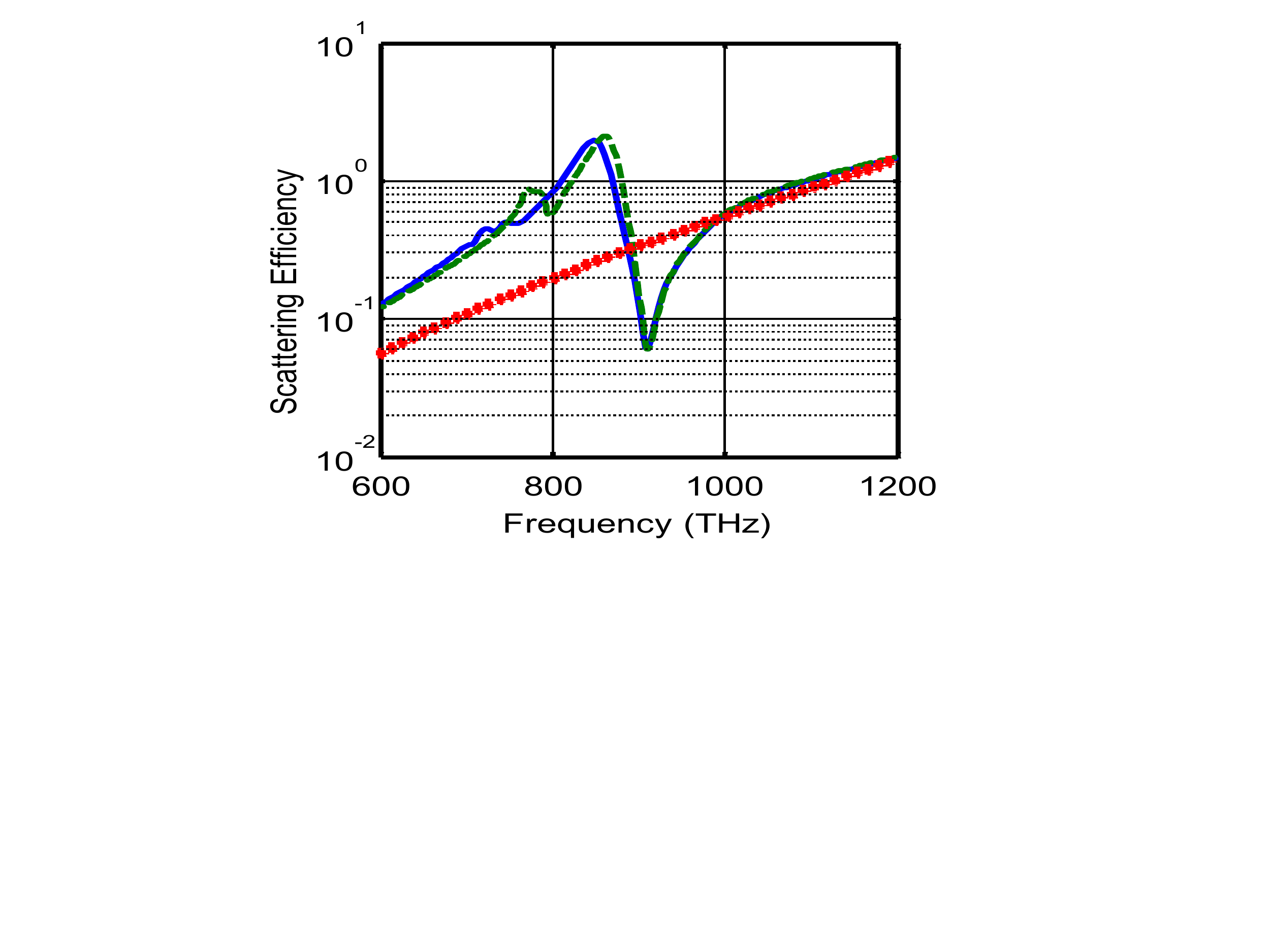}}
\caption{Numerical calculation of the total scattering efficiency for two different permittivities of the  dielectric core sphere as a function of frequency
[(a)  $\varepsilon_\mathrm{c}~=~2.25$, (b) $\varepsilon_\mathrm{c}~=~8$]. The lines have then following meaning: red dotted line - bare sphere; blue solid line - core-shell system rigorously calculated where the fine details of the structure are accounted for; green dashed line - homogeneous core - shell system where the shell is described by the Maxwell-Garnett approximation (Eq. \ref{mg_eq}). Note the very good agreement in the frequency domain of effective cloaking around 900 THz.}
\label{scs1}
\end{center}
\end{figure}
\begin{figure}[h!]
\begin{center}
{\includegraphics[width=8cm,angle=0]{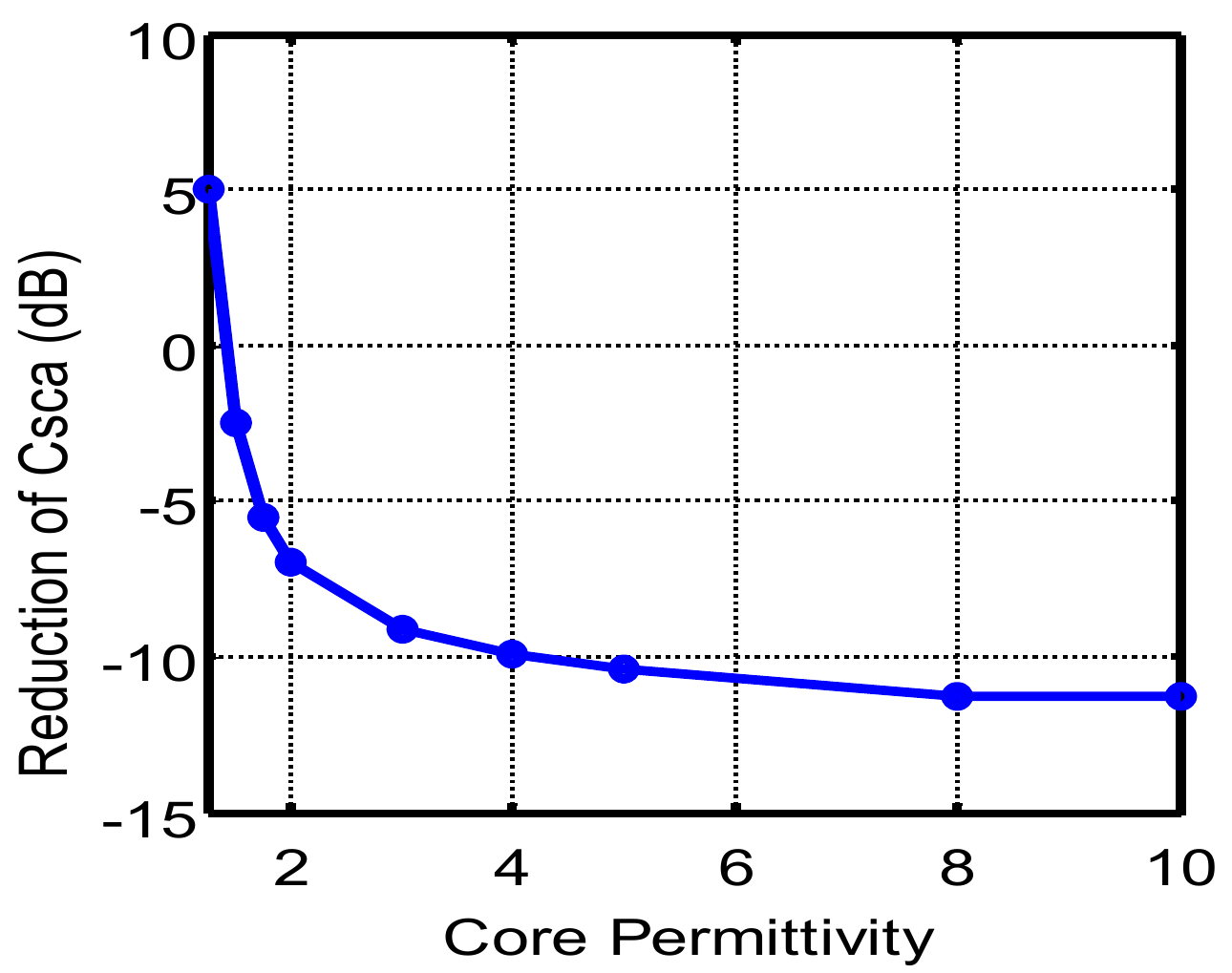}}
\caption{Efficiency of the cloaking mechanism versus the dielectric function of the core sphere in dB. It shows that for high permittivities core spheres the reduction of scattering is more than 75 percent.}
\label{cloak_efficiency}
\end{center}
\end{figure}
\begin{figure}[h!]
\begin{center}
%\hspace{-1.25cm}
{\bf (a)} \scalebox{0.325}{\includegraphics{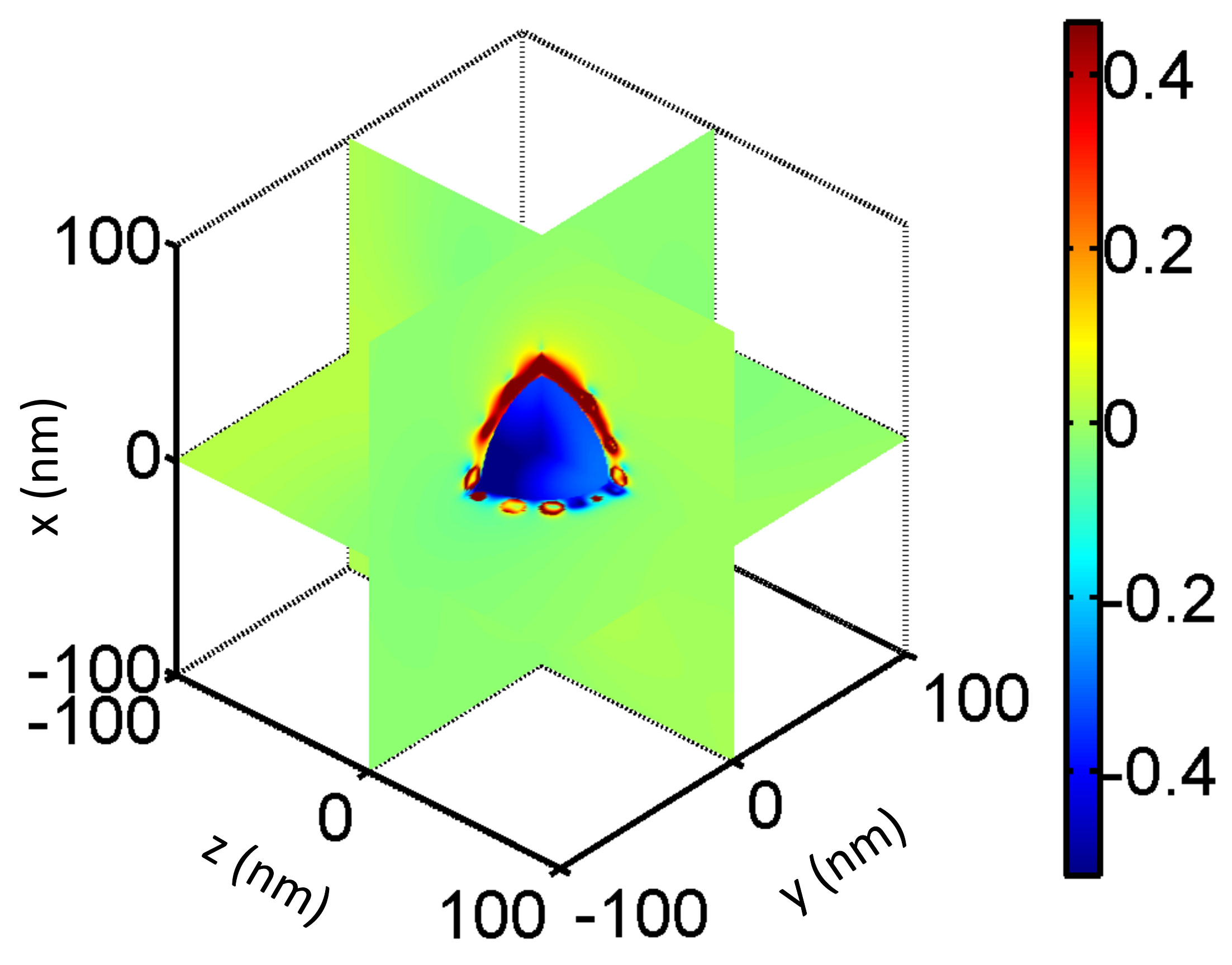}}
%\vspace{1cm}
%\hspace{-0.55cm}
{\bf (b)} \scalebox{0.325}{\includegraphics{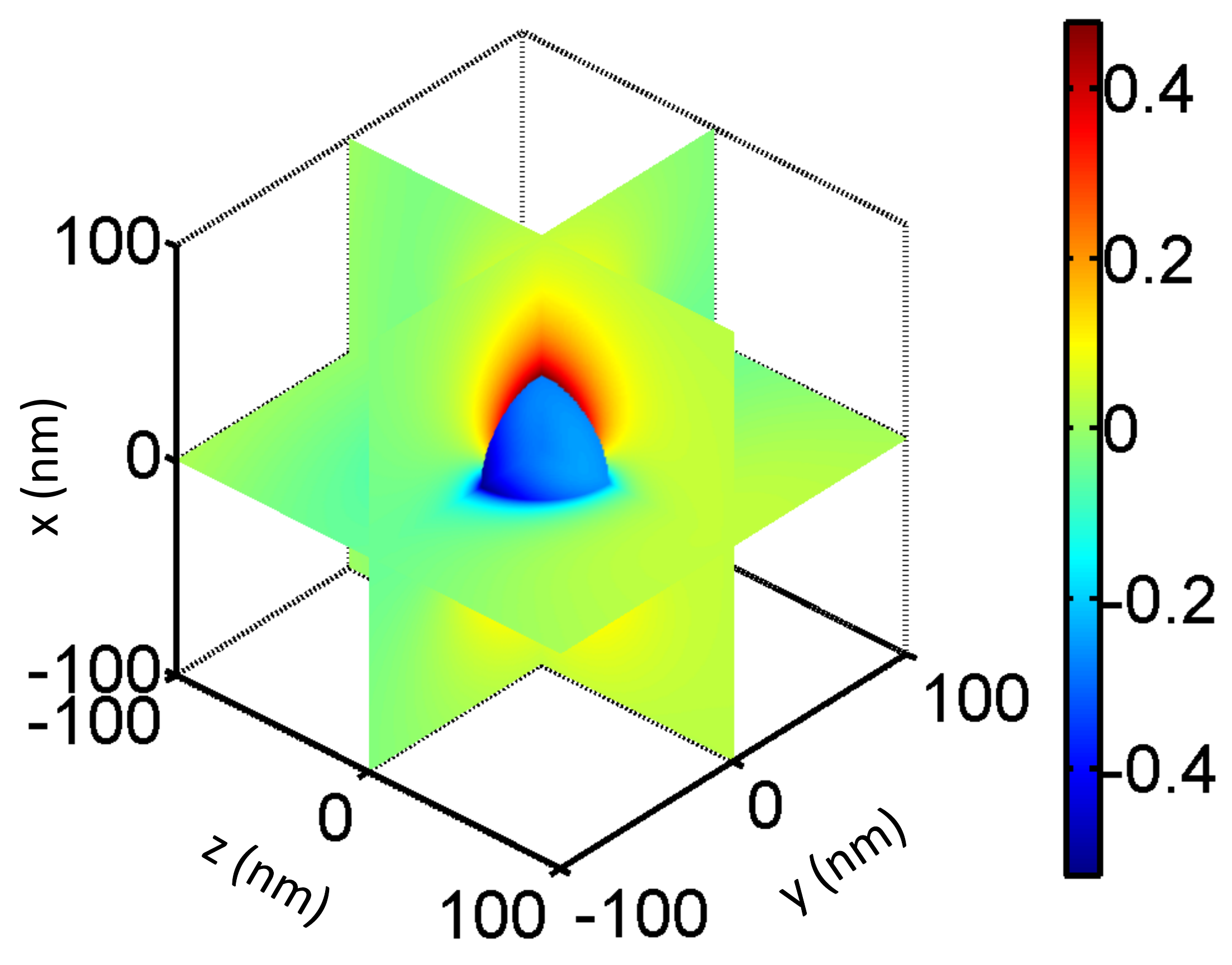}}
\caption{Time averaged field distributions in a logarithmic scale of a dielectric sphere of $\varepsilon_c=8$ which is cloaked by the NPs (a) and on its own for comparison (b). The structures are illuminated with a unit amplitude plane wave (909.5 THz) propagating in the $z$-direction and being polarized in the $x$-direction.}
\label{fieldmap}
\end{center}
\end{figure}
In what follows, we aim at numerically evaluating the performance of the cloak. As it  is commonly known, the electromagnetic scattering cross section (SCS) per unit length of an arbitrarily object (given by Eq. \ref{scs}) is a quantitative measure of its visibility. In Fig. \ref{scs1} the response of the cloak is displayed  for the frequency range between 600 and 1200 THz. The scattering efficiency of the cloak is defined as the ratio of the SCS of the covered object normalized to its geometrical cross section. It was calculated by two different methods. The first method is rigorous by relying on a multiple scattering formalism for a large number of spheres \cite{muhlig}. The object under consideration was a large dielectric core sphere (radius $35$ nm) with  131 small silver NPs (radius $5$ nm) amorphously distributed with a filling fraction of  $f$ = 0.34. A schematic  of such a structure is shown in Fig.~\ref{scheme}. This core-shell system was illuminated by a linearly polarized plane wave. Due to the spherical symmetry of the object and the amorphous nature of the shell the response is qualitatively fully preserved for various implementations of nominally the same geometry. In a second simulation the scattering response of the core-shell system was calculated where the shell was described as an effectively homogeneous medium with a permittivity according to the Maxwell-Garnett formula (\ref{mg_eq}). As a reference the scattering efficiency for the uncloaked particle is shown too. Two different permittivities of the core were considered ($\varepsilon_\mathrm{c}~=~2.25$ and $\varepsilon_\mathrm{c}~=~8$). It can be clearly seen that in a finite frequency range an excellent scattering reduction may be achieved. The frequencies correspond to the domain where the effective permittivity of the shell takes the required small and positive value. The scattering efficiency of the particle to be cloaked is reduced by approximately 70\% and the cloaking behavior is equally well predicted by the approximate treatment. In contrast, cloaking is inefficient for the second possible case, i.e. for a large, but negative effective permittivity. This second operational frequency can be disclosed from the analytical treatment to be either 830 THz for $\varepsilon_\mathrm{c}~=~2.25$ or 790 THz for $\varepsilon_\mathrm{c}~=~8$. It is noteworthy that rigorous simulations do not show a peculiar behavior at these frequencies. This is a clear indication that near NP resonance the description of the shell in terms of an effective permittivity fails since the impinging light probes explicitly all metallic NPs. This can also be  recognized by looking at the rigorously calculated spectra that feature many small resonances in this domain. Moreover, as anticipated before, this solution to the cloak is inefficient because unavoidably absorption comes into the play.

Nevertheless, despite all these details, it can safely be concluded that the cloak operates as anticipated and the remaining deviations can be explained by considering the peculiar details of its implementation. It remains interesting to note, that the higher the permittivity of the obstacle (and thus its 'dielectric size') the better the efficiency of the shell. To summarize the cloaking efficiency we also provide the scattering reduction versus the permittivity of the core in Fig.~\ref{cloak_efficiency} confirming that for large permittivities the reduction of scattering is more than 75 percent. To further check the functionality of the cloak, we show in Fig.~\ref{fieldmap} the amplitude distribution of the electromagnetic field scattered by the spherical obstacle with [Fig.~\ref{fieldmap} (a)] and without the plasmonic shell ([Fig.~\ref{fieldmap} (b)]). When it is surrounded by the cloak,  the scattered amplitude is close to zero everywhere in space in contrast to the uncloaked case. As already explained the reduction of scattering is due to the proper choice of the permittivity function of the plasmonic cover. This is consistent with the scattering reduction predicted in Fig.~\ref{scs1}. It finally remains to mention that the same mechanism is also expected to work for metallic core particles with finite or infinite conductivity where, however, the required specific effective shell permittivities will be different.

\section{Conclusion}

In the current paper, we propose a realistic design for cloaking three-dimensional objects by suppressing their scattering response in the dipolar limit. This technique is based on the well-known plasmonic cloaking which relies on the use of low permittivity shells. We propose to extend this idea by allowing the core sphere to be covered by a finite number of small metallic nanoparticles. In the dipolar limit this medium can be described as effectively homogenous possessing a Lorentzian resonance in the effective permittivity. The description of this medium is feasible by using the Maxwell-Garnett approximation which allows for an analytical description of the cloak in the dipolar limit. We even went one step further and compared the analytical design to full-wave simulations of the entire core-shell system by using a multiple scattering algorithm that explicitly takes into account all the individual spheres forming the shell. A scattering reduction of approximately 70 \% has been proven.  Using this design one may envision that  cloaking theory can be moved closer to its practical and feasible realization for optical waves.
We are confident that our proposed cloak can be implemented in an optical experiment where bottom-up methods from colloidal nanochemistry are used to decorate a dielectric sphere by sufficiently densely packed small metallic nanoparticles. Moreover, we are working towards the extension of this technique to cloak larger obstacles for which higher order multipolar contributions to the scattered field have to be taken into account. For such objects the effective permeability of the shell equally needs to be controllable for which various unit cells that rely on metallic NPs can be envisioned.

\section*{Acknowledgments}
Financial support by the Federal Ministry of Education and Research PhoNa, from the State of Thuringia within the ProExcellence program MeMa, as well as from the European Union FP7 project NANOGOLD is acknowledged.

\end{document}